\documentclass[sigconf]{acmart}

\usepackage{booktabs} 
\usepackage{xurl} 
\usepackage{amsfonts}
\usepackage{tabularx}
\usepackage{siunitx}
\usepackage{enumitem}
\usepackage{stfloats}
\usepackage{geometry}
\usepackage{multirow}
\usepackage{makecell}
\usepackage{tcolorbox}
\usepackage{stackengine,graphicx}
\usepackage{xspace}
\usepackage{caption}
\usepackage{subcaption}
\usepackage{graphicx}
\usepackage{makecell}
\usepackage{multirow}
\usepackage{pgf}
\usepackage{url}
\usepackage{color}
\newcounter{querycount}
\makeatletter
\newcommand{\myitem}[1]{%
\item[#1]\protected@edef\@currentlabel{#1}%
}
\makeatother

\copyrightyear{2024}
\acmYear{2024}
\setcopyright{rightsretained}
\acmConference[CIKM '24]{Proceedings of the 33rd ACM International Conference on Information and Knowledge Management}{October 21--25, 2024}{Boise, ID, USA}
\acmBooktitle{Proceedings of the 33rd ACM International Conference on Information and Knowledge Management (CIKM '24), October 21--25, 2024, Boise, ID, USA}
\acmDOI{10.1145/3627673.3680268}
\acmISBN{979-8-4007-0436-9/24/10}

\begin{document}

\title{Leveraging Knowledge Graphs and LLMs to Support and Monitor Legislative Systems}

\author{Andrea Colombo}
\authornote{Supervised by prof. Stefano Ceri (Politecnico di Milano)}
\affiliation{%
  \department{Dipartimento di Elettronica, Informazione e Bioingegneria,}
  \institution{Politecnico di Milano}
  \city{Milan}
  \country{Italy}
}
\email{andrea1.colombo@polimi.it}


\begin{abstract}
Knowledge Graphs (KGs) have been used to organize large datasets into structured, interconnected information, enhancing data analytics across various fields. In the legislative context, one potential natural application of KGs is modeling the intricate set of interconnections that link laws and their articles with each other and the broader legislative context.

At the same time, the rise of large language models (LLMs) such as GPT has opened new opportunities in legal applications, such as text generation and document drafting. Despite their potential, the use of LLMs in legislative contexts is critical since it requires the absence of hallucinations and reliance on up-to-date information, as new laws are published on a daily basis.

This work investigates how Legislative Knowledge Graphs and LLMs can synergize and support legislative processes. We address three key questions: the benefits of using KGs for legislative systems, how LLM can support legislative activities by ensuring an accurate output, and how we can allow non-technical users to use such technologies in their activities. To this aim, we develop Legis AI Platform, an interactive platform focused on Italian legislation that enhances the possibility of conducting legislative analysis and that aims to support lawmaking activities.
\end{abstract}




\maketitle

\section{Introduction}
Knowledge Graphs (KGs) have been widely popular in recent years due to their ability to organize vast amounts of complex data in a way that is easily navigable and semantically rich. KGs have found applications across diverse fields, from healthcare and finance~\cite{liu2019anticipating} to biology~\cite{choi2019inference} and education~\cite{sun2016visualization}. Their power lies in transforming unstructured data into structured, interconnected information, enabling more sophisticated data analytics~\cite{lu2017analysis,hodler2022graph}, for instance, by utilizing graph algorithms~\cite{graphAlgorithms} or by improved machine learning outcomes and more intuitive user interactions. 
In the context of legislative systems, a Knowledge Graph model can be used to represent legislative data that interconnects various entities such as bills, statutes and legislators. It enables an easy integration and retrieval of complex legislative information, providing insights into the relationships and dependencies between different components of the legislative framework. 
Given the unstructured nature of legislative texts, the main challenge in this context is to build a domain-specific Knowledge Graph, i.e., an accurate and reliable KG~\cite{LLMKG}, that could be used to conduct interactive and quantitative analysis. 

More recently, with the advent of large language models (LLMs), such as GPT or LLama, many practical applications of LLMs in the legal context are possible, such as legal text classification, information extraction and retrieval pipelines that have started to appear~\cite{SANSONE2022101967,legalWordEmbedding}, with also some first attempt in more critical tasks such as Legal Documents Drafting~\cite{lam2023applying}. However, in the latter case, the presence of hallucinations makes the task much harder due to the critical nature of the drafting of a new legal document, be it a new law or a contract. 
LLMs are, in fact, black-box models that fall short of capturing and accessing factual and updated knowledge, which undermines the results of such tools, for instance, by not presenting a proper legal foundation for drafted legal documents. In contrast, Knowledge Graphs (KGs) store up-to-date external knowledge but do not offer users the friendly access and human interpretability that an LLM can achieve.

In this work, we analyze how we can use Legislative Knowledge Graphs and LLM to support legislative processes. In particular, our research questions might be summarized as follows:

\begin{itemize}
    \addtocounter{querycount}{1} \myitem{RQ\thequerycount} \label{rq1} \textit{Which are the benefits of using Knowledge Graphs to represent legislative systems?}
    \addtocounter{querycount}{1} \myitem{RQ\thequerycount} \label{rq2}
    \textit{How can we use Large Language Models to support legislative activities in an accurate manner and without hallucinations?}
    \addtocounter{querycount}{1} \myitem{RQ\thequerycount} \label{rq3}
    \textit{How can we present the result of these approaches to interested stakeholders, mainly non-IT users?}
\end{itemize}

To answer such research questions, we have been inspired by recent literature showing the benefits of leveraging the synergy between a domain-specific Legislative Knowledge Graph - that we recently introduced in ~\cite{ilpgCikm} - and LLMs to present useful applications of knowledge graphs and large language models applied to the legislative context, with first attempts to combine the best of the two worlds, via the use of so-called GraphRAG approaches~\cite{LLMKG}. To this aim, we focused on the Italian legislation and we developed a multi-purpose, user-friendly platform that directly communicates with the KG, allowing users and interested stakeholders to (i) analyze the quality of enacted laws stored in the graph via linguistic metrics and LLMs, (ii) receive support for the drafting of new laws, (iii) interactively analyze the quality and the complexity of the overall legislative system on a temporal dimension.


\section{Background}

The foundations of this work rely on the possibility of representing legislative systems in a Knowledge Graph. While most of the literature has notably focused on RDF-based graph approaches~\cite{anelli2022navigating,angelidis2018a}, pushed by the Semantic Web community and the need to link knowledge bases by providing globally unique and resolvable identifiers over multiple domains, we looked at a more flexible solution, which could offer more graph traversal capabilities and a more compact representation of the complexities of legislative systems. Our proposed solution was to use property graphs, a type of data structure where nodes (vertices) and edges (relationships) can have associated properties (key-value pairs)~\cite{angles2018property}. This allows for detailed and complex information to be captured within the graph, facilitating advanced querying and analysis. 
In addition, the recent development of expressive conceptual, machine-readable XML models of various aspects of general legislative knowledge, such as the Legal Knowledge Interchange Format (LKIF)~\cite{hoekstra2007lkif}, LegalRuleML~\cite{legalRuleML} and Akoma Ntoso (AKN)~\cite{akomantoso} simplify the possibility of transforming an initially unstructured source of data into a graph database. In particular, AKN has also been adopted by a wide range of international organizations and national legislation, such as the European Parliament~\cite{EUAKN} or the Italian bureaucracy, making it also useful for developing a consistent, interchangeable pipeline that is applicable to distinct legislative systems. In fact, by leveraging the XML tags of the standard, we can gather knowledge graphs consistent across national legislations, making them comparable. 

\smallskip \noindent \textbf{The Italian Legislative Property Graph} In ~\cite{ilpgCikm} we presented both the schema of a Legislative Property Graph and we applied it to the Italian legislation, whose laws are available in AKN. 
This also included an ETL pipeline from Normattiva~\cite{normattiva}, the official data source of Italian laws, to extract relevant properties of nodes and edges, enriching the overall model. These include the domain of reference identified by the ministry that signed the law or the specificity of the references, i.e., whether the reference further specifies a paragraph of interest.

\section{Overview of the Legis AI Platform}
\begin{figure}
    \centering
    \includegraphics[width = \linewidth]{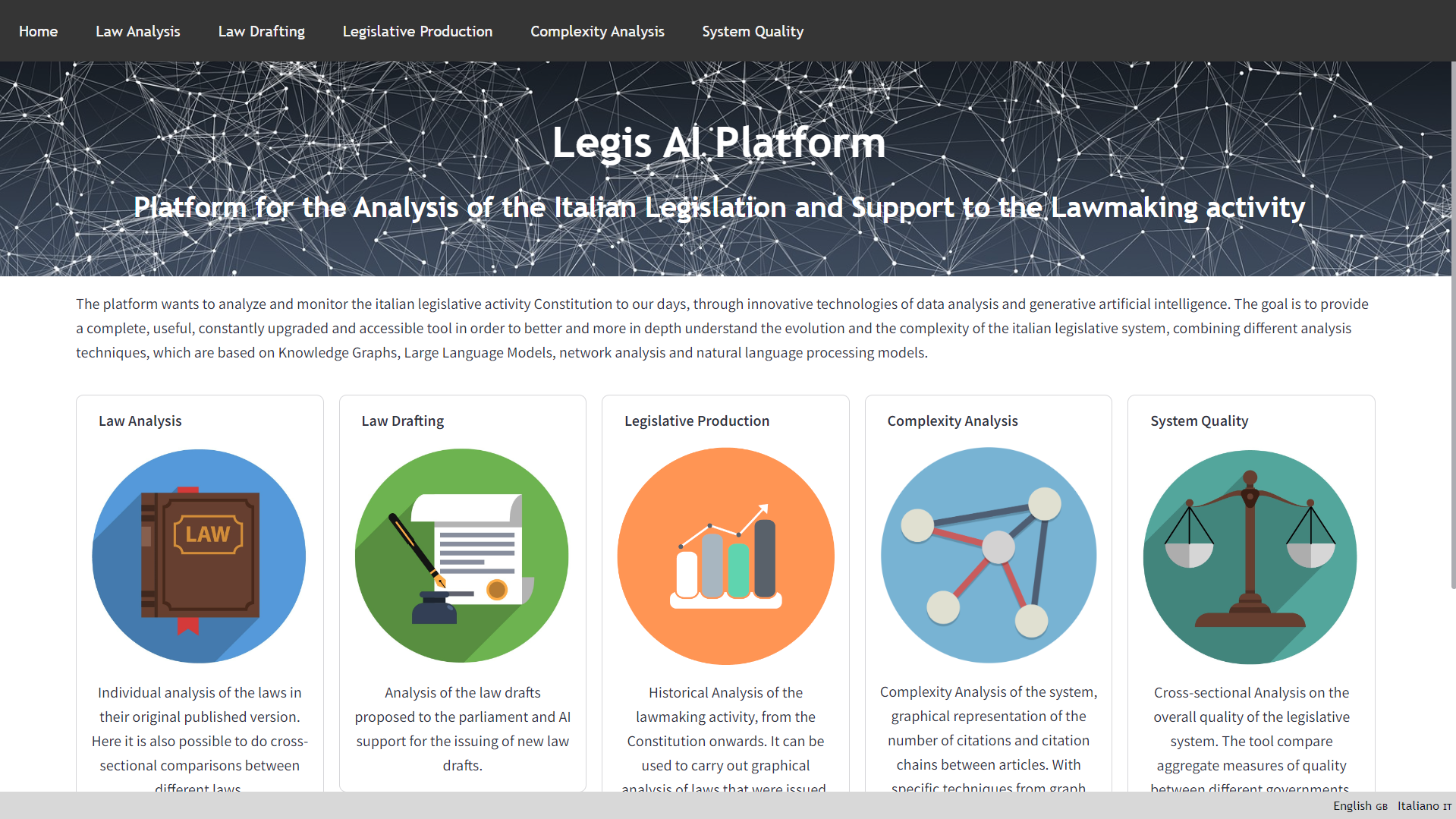}
    \caption{Home Page of the Legis AI Platform, allowing access to all the modules implemented.}
    \label{fig:homePage}
    \vspace{-2mm}
\end{figure}

The Legis AI Platform platform aims to analyze, monitor, and support Italian legislative activity by utilizing graph database technologies and artificial intelligence, with integrations that leverage the capabilities of generative AI in a precise and effective manner. The goal is to provide a comprehensive, real-time, updated, and easily accessible analysis tool to better understand the evolution and complexity of the Italian legislative system.
Its home page can be seen in Figure~\ref{fig:homePage}, while the platform can be accessed at \href{http://gmql.eu/legisplatform}{Legis AI Platform}.
In the next sections, we present its main modules, with a particular focus on the underlying approaches that have been developed and implemented in the front-end.

\subsection{Law Analysis}
\label{sec:lawAnalysis}
The first module of the platform focuses on the individual analysis of legal texts by illustrating their metadata, querying the Legislative Knowledge Graph, and employing linguistic metric functions to derive insights into the textual quality of the law. All of this is then presented to the user in a readable textual report, obtained by leveraging the capabilities of LLMs to create fluent text starting from a list of factual inputs.
Here, we leverage the KG as a plain data source to retrieve the laws of interest. For each law, linguistic metrics are computed and presented both in the form of indicators and within a report that is generated on the fly by an LLM and presented in the UI. In this module, we try to tackle \ref{rq2} and \ref{rq3} from a practical perspective: by limiting the task of the LLM to the generation of textual reports and providing all the necessary input, we manage to leverage the enormous potential in generating text but without incurring in major distortion within the text and presenting a quality report in a user-friendly and interactive format, i.e., allowing users to choose the set of laws to be compared with. 


\smallskip \noindent \textbf{Textual Analysis}. After the selection of a law of interest, linguistic quality metrics are computed. In particular, we consider simple metrics such as the average length of words, the average length of sentences, the relative number of gerunds, adjectives, and pronouns, and more complex readability indexes that are popular in the linguistic computing literature, namely, the Flesch Reading Ease index, the Embedding and Center Embedding indices and the Gulpease index, which is specifically designed for the Italian language. All of them are good indicators of textual complexities in laws~\cite{martinez2022so}.


\smallskip \noindent \textbf{Report Generation}. The readability and comprehensibility indices are used to generate textual reports using generative artificial intelligence. Users can select a set of laws of interest against which a specific law can be compared. For instance, you can compare the quality of a newly published law against the set of laws referring to the same topic. As the model is specifically parametrized and requested not to provide recommendations, the model is not subject to the problem of producing "hallucinations" and sticks to the task of presenting the findings and the result of the analysis in a readable form. To this aim, we use a LLama-3 70B model~\cite{llama}, combined with prompt engineering to respond to the task of report generation, i.e., by providing an example of the output.

\subsection{Supporting the Lawmaking Process}
The second module of the platform addresses other aspects of all research questions as it aims to assist users in the legislative initiative process in two ways: by allowing a preventive analysis of the laws currently under review - such that their text can be improved before publication - and by providing support for the identification of the legislative landscape regarding a certain topic the legislator might want to address.

\smallskip \noindent \textbf{Preventive Analysis of Draft Laws}.
Each proposal that is in discussion or has been presented in the Italian parliament can be gathered through a dedicated API~\footnote{\url{https://dati.camera.it/sparql}}. For such soon-to-be laws, metadata are extracted and presented to users, such as the proponent's name, which is then linked to its biographical information via the connection to Wikidata. 
In this module, we also implemented the functionality we envisioned for Section~\ref{sec:lawAnalysis}. However, in this scenario, we are allowing a preventive analysis, which could help the legislative initiative process be driven towards a better quality of the proposals, bringing in less ambiguity and complexity in the text.


Unlike the comparison proposed for existing laws, here the task is more complex. In fact, while for the latter, users could choose set of laws based on pre-defined metadata of the law, such as the publication year or the domain, their absence in drafted laws raises the problem of defining a set of laws to be used for this task. 
To this end, we developed an approach that first uses LLMs to identify the relevant laws that treat the same topics as the draft ones. This approach can be summarized by the following three steps:
\begin{enumerate}
    \item \textbf{Topic Extraction}. First, an appropriately parameterized LLama-3 70B model is asked to identify the reference topics for the title or keywords entered. For example, if "regulation of artificial intelligence technologies" is entered, the corresponding domains will be "computer science" and "innovative technologies". 
    \item \textbf{Extraction of Relevant Laws via Textual Embedding}. Using a textual embedding model~\footnote{text-embedding-3-small}, \textit{in-force} laws~\footnote{As demonstrated in~\cite{ilpgCikm}, the derivation of in-force laws only can be easily achieved via the Property Graph model} that address the same topics are retrieved. For this purpose, a Vector Index on the Knowledge Graph is used, enabling an efficient search called Approximate Nearest Neighbor (ANN) through the Hierarchical Navigable Small World (HNSW) algorithm.
    \item \textbf{Generation of the Report}. As for the analysis of existing laws, the LLama-3 70B model is asked to generate the final report based on the statistics calculated on the new bill proposal.
\end{enumerate}

\smallskip \noindent \textbf{Identification of the Normative Landscape}.
In a related section of this module, we also developed an approach to support the drafting of new law proposals, allowing the identification of the normative framework within which they will be inserted. To this end, we resort to a Knowledge Graph Retrieval Augmented Generation-like (GraphRAG) approach, a powerful way to combine graphs and LLMs to improve the outputs of the latter. The idea here is to detect which articles or laws might be useful or that provide the legal foundations for a potential new law proposal. In other words, it deals with detecting the references that are relevant if I would like to legislate on a certain topic.
We adopted such an approach in this context, considering the requirement to reduce the risk of hallucinations. This resulted in the following pipeline:
\begin{enumerate}
    \item \textbf{Text Input}. Through a textual prompt in the front-end, users can input a potential title or a series of keywords that describe the proposal they would like to investigate.
    \item \textbf{Topic Extraction}, as in the previous section, although here it is performed on the user input.
    \item \textbf{Topic Expansion}. As the input text from the users might be rather vague, we added a step of topic expansion. In this step, we send our LLM - a LLama-3 70B model -the keywords extracted and the textual input and we ask it to derive close topics that might relate to the ones indicated by the user.
    \item \textbf{Extraction of Relevant Laws via Textual Embedding}, as in the previous section. The set of laws resulting from this step might be used by the user to understand the details of the legislative landscape within which the new law will be inserted.
    \item \textbf{Query the Knowledge Graph}. Finally, we navigate the Knowledge Graph to query the articles or laws that are most frequently used in the set of relevant laws detected in (4). Here, the granularity of our KG, which distinguishes between types of references in laws, allows us to focus specifically on \textit{preamble} citations, which are the legal foundation for a law.
\end{enumerate}

\begin{figure}
    \centering
    \includegraphics[width = \linewidth]{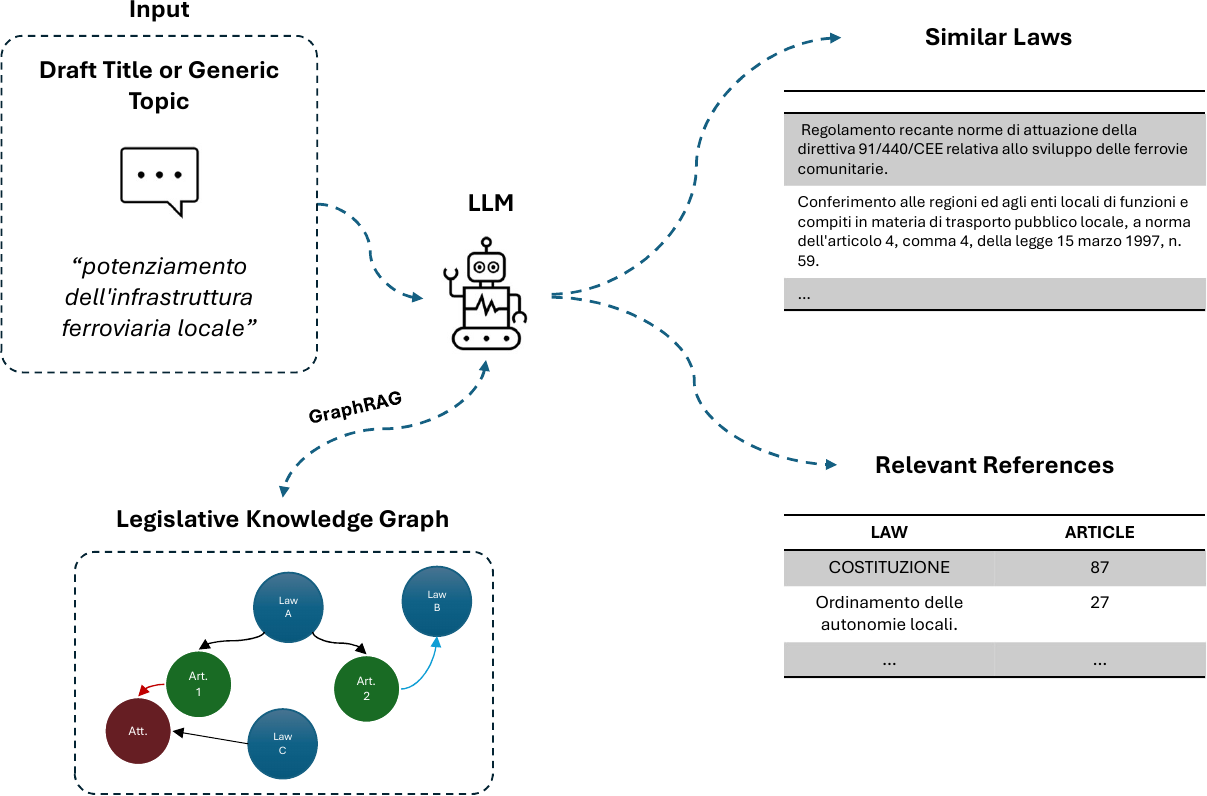}
    \caption{GraphRAG workflow for the draft of new laws}
    \label{fig:lawDraft}
    \vspace{-3mm}
\end{figure}

Figure~\ref{fig:lawDraft} depicts an instantiated overview of this approach.
The result is twofold. First, relevant similar laws that could interest the new proposal are suggested to the user, supporting the retrieval of the legislative landscape. Then, relevant references are reported: such laws or articles are ordered by the relative frequency, providing an indication of how likely such articles should be cited to provide a legal foundation for the new law. This set helps in speeding up the lawmaking activities, since it reduces the number of background work required to retrieve the laws and the articles that might be influenced by the new proposal.

\subsection{Monitor System Complexity}
In the remaining modules, we use the Knowledge Graph representation to perform interactive analyses, mainly on an aggregate level. Specifically, we let the user perform queries over the legislative system by leveraging a user-friendly front-end that communicates with the graph through Cypher, Neo4j's declarative query language. Users can modify parameters, such as the temporal granularity of interest, and download the resulting datasets from the analyses. We already showed how a graph data model is very good at representing the complexity of the interconnection of a legislative system~\cite{ilpgCikm}, allowing non-trivial queries that compute system complexity metrics or that allow users to monitor its temporal evolution.

\section{Evaluation Approach and Future Work}
We plan to conduct two studies to validate the results and metrics derived from our approaches and presented in our platform. First, we will perform a user study, asking legal experts (e.g., law students) to evaluate the quality of a sample of laws or even individual law articles. The results of this study will allow us to validate the indices presented on the platform and the reports generated by the LLM. Then, we will validate the second module of Legis AI Platform by conducting an empirical study: we will use recent laws that have come into force as a test set to calculate recall and precision measures, determining the tool's usefulness in identifying relevant acts that have been in fact used as the legal foundation of the law.

In addition, we plan to continue our interactions with the potential users of the platform, namely legislator representatives (i.e., Camera dei Deputati), journalists, and researchers who have expressed their interest in such a tool. More specifically, we will try to understand how to leverage the graph representation to perform network analysis, eventually by combining them with LLMs.

In future iterations, we plan to further explore LLM's potential synergies with Legislative Knowledge Graphs. In particular, we plan to develop interactive ChatBots that can leverage the data model either to answer complexity queries made in natural language, by using Text-to-Cypher approaches, or by using it as an assistant that helps investigate the content of the laws and using the inter-dependencies of the KG to provide more complete responses. While we are now focusing on the Italian landscape, we also plan to replicate or even compare distinct national legislation with a project over the US federal laws that has already started.

\section{Conclusion}
The platform was built to assist in the supervision and support of legislative processes. This was achieved by developing a user-friendly platform based on the most recent technologies and approaches adapted to the context of computer law. We focused on the reporting activity and on supporting the lawmakers in drafting new laws. Great care has been taken to ensure the absence of "hallucinations" and to maintain a neutral stance of the AI models, thus preventing them from providing recommendations or opinions on the issues. This was achieved through a combined approach of prompt engineering, retrieval-augmented generation on a graph, and objective metrics for evaluating the quality of laws.

\smallskip \noindent \textbf{Resources} The platform is available at \url{http://gmql.eu/legisplatform/?lang=en}

\smallskip \noindent \textbf{Ackowledgments}
Andrea Colombo kindly acknowledges INPS for funding his Ph.D. program, thanks Claudio Michelacci and Luigi Guiso from EIEF for the fruitful discussions and the Applied Research Team from Banca d'Italia for the support.

\bibliographystyle{ACM-Reference-Format}
\bibliography{biblio}

\end{document}